\documentclass[prl,twocolumn,showkeys,showpacs,preprintnumbers,amsmath,amssymb]{revtex4} 
\usepackage{graphicx}
\usepackage{dcolumn}
\usepackage{bm}
\usepackage[tight]{subfigure}
\usepackage{enumerate}
\usepackage{amsmath}
\usepackage{verbatim}
\usepackage{color}
\usepackage{natbib}
\usepackage{float}

\begin{document}
\title{Tuning the two-electron hybridization and spin states in parallel-coupled InAs quantum dots}

\today
\author{Malin Nilsson$^{1}$}
\author{Florinda Vi\~nas Bostr\"om$^{1}$}
\author{Sebastian Lehmann$^{1}$}
\author{Kimberly A. Dick$^{1,2}$}
\author{Martin Leijnse$^{1}$}
\author{Claes Thelander$^{1}$}

\affiliation{$^{1}$Division of Solid State Physics and NanoLund, Lund University, Box 118, S-221 00 Lund, Sweden}
\affiliation{$^{2}$Center for Analysis and Synthesis, Lund University, Box 124, S-221 00 Lund, Sweden}

\begin{abstract}
We study spin transport in the one- and two-electron regimes of parallel-coupled double quantum dots (DQDs). The DQDs are formed in InAs nanowires by a combination of crystal-phase engineering and electrostatic gating, with an interdot tunnel coupling ($t$) tunable by one order of magnitude. Large single-particle energy separations (up to 10 meV) and $|g^*|$ factors ($\sim$10) enable detailed studies of the $B$-field-induced transition from a singlet-to-triplet ground state as a function of 
$t$. In particular, we investigate how the magnitude of the spin-orbit-induced singlet-triplet anticrossing depends on $t$. For cases of strong coupling, we find values of 
230~$\mu$eV for the anticrossing using excited-state spectroscopy. Experimental results are reproduced by calculations based on rate equations and a DQD model including a single orbital in each dot.

\end{abstract}

 \maketitle

Coupled spin states as a base for qubits is today a widely
studied topic in research on quantum information processing\cite{Loss1998,Petta2005,Nadj-Perge2010B}. A standard approach to forming a two-spin qubit is to manipulate entangled singlet and triplet states
in a semiconductor-based double quantum dot (DQD). Materials with significant spin-orbit interaction (SOI), such as InAs, are of particular interest, because spin states
can be manipulated by inducing small changes in the electron orbitals \cite{Nadj-Perge2010B}. The need for precise spin control has triggered studies of spin dynamics in DQDs in a pursuit to
improve the understanding of such spin-nonconserving interactions \cite{Pfund2007,Nadj-Perge2010}.
 
In serial-coupled DQDs, spin relaxation processes are generally probed via the leakage current in the Pauli spin blockade regime \cite{Johnson2006, Pfund2007,Pfund2007B,Danon2009,Nadj-Perge2010}. There, transport is studied near the degeneracy point of the (2,0) and (1,1) electron occupation
in the DQD, where the leakage current is a result of the coupling between singlet $S$(0,2) and triplet $T$(1,1) states. DQDs that instead are parallel coupled to source and drain
have no such blockade mechanism, but open up for direct studies of the first-order electron-spin transitions by dc transport measurements. Extensive studies have been conducted on GaAs parallel-coupled DQDs by Hatano \textit{et al.} \cite{Hatano2005,Hatano2008,Hatano2011,Hatano2013}.
Here,  a small $|g^*|$ factor was limiting the resolution in the measurements and, due to the weak effective SOI in GaAs, no SOI-mediated anticrossing of $S$(1,1) and $T$(1,1) states was reported.  

In GaAs, hyperfine interaction has been identified as an important contribution to spin relaxations \cite{Khaetskii2000,Khaetskii2001,Koppens2005}.  However, in InAs, the relative contribution of hyperfine interaction is small in the strong interdot tunnel coupling regime \cite{Danon2009,Nadj-Perge2010}.
Here,  the coupling between  singlet $S$(1,1) and triplet $T$(1,1) can be directly evaluated by the
magnitude of the avoided crossing when the states are
aligned by an external magnetic field, similar to what has
been reported for single quantum dots (QDs)  \cite{Fasth2007,Nilsson2009}.

In this Letter, we present a highly controlled, yet simple to fabricate, system of fully depletable parallel-coupled quantum dots with widely tunable interdot tunnel coupling ($t$). Large values for the $|g^*|$ factor ($\sim$10), single-QD orbital energy spacing (up to 10~meV) and intradot charging energy ($>$10~meV) provide a wide window for fundamental studies of electron-spin properties. We focus here on the transition between one and two electrons (1$e$/2$e$) involving the $S$(1,1) and $T$(1,1) states, where one electron populates each dot. In particular, we can clearly resolve the magneticfield-induced evolution of the excited-states spectrum and demonstrate $B$-field-dependent tuning of the   2$e$ ground state from singlet $S$(1,1) to triplet $T_+$(1,1). We show that the magnitude of the SOI-induced anticrossing of  $S$(1,1) and $T_+$(1,1) states can be tuned over a wide range with $t$, up to $\Delta^*_{ST}$ = 230 $\mu$eV for $t$ = 2.3 meV.
The experimental results are reproduced by a DQD model with a single level on each dot, including exchange energy, spin-orbit coupling, and Zeeman energy.

Parallel-coupled DQDs are formed by a combination of hard-wall tunnel barriers to source and drain and a set of three gate electrodes \cite{Nilsson2017}. First, a single QD is formed
by controlling the crystal phase in InAs nanowires. Thin segments of wurtzite (WZ) crystal phase in the otherwise zinc blende (ZB) nanowires act as hard-wall tunnel barriers
for electrons \cite{Dick2010,Nilsson2016A,Nilsson2016B,Chen2017}. Source, drain, and side-gate contacts (Ni=Au 20=80 nm) are fabricated as detailed in Ref.~\cite{Nilsson2017}, whereas the substrate acts as a global back-gate. 
All measurements are performed in a dilution refrigerator at an electron temperature of $\sim$50 mK. When applied, the external $B$ field is aligned perpendicular to the substrate and
points close to a $\langle 112 \rangle$-type crystal direction of the nanowire (see the Supplemental Material). 

\begin{figure}[t]
\centering
\includegraphics[width=\columnwidth]{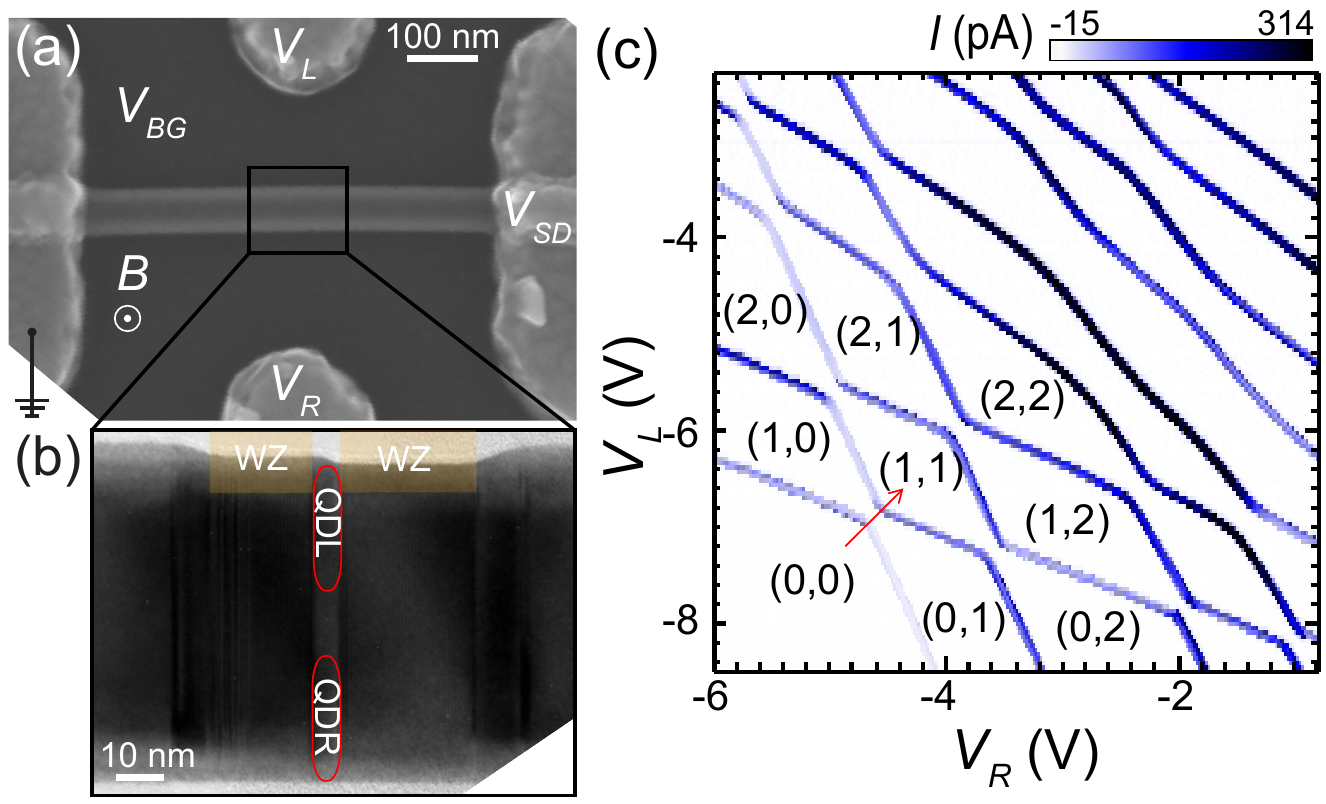}
\caption{(Color online) ((a) A scanning electron microscope image of a typical
DQD device where the three gates (left and right side gate and global back gate) and source and drain electrodes are indicated. (b) High-resolution transmission electron microscopy image
viewed along a $\langle 110 \rangle$-type direction of the disc-shaped ZB QD (4~nm) sandwiched between WZ (22 and 28~nm) segments (highlighted). The diameter of the single QD is 67~nm. The
positions of the left and right QDs, in a simplified picture, are indicated. (c) Charge stability diagram where the electron populations of the two QDs are indicated ($V_{BG}=0$~V, $V_{SD}=1$~mV) [cooldown I]. The red arrow indicates the gate vector used in Fig.~\ref{fig2}. }
\label{fig1}
\end{figure}

A scanning electron microscope image of a typical device and a transmission electron microscopy image of a crystalphase QD are shown in Figs.~\ref{fig1}(a) and \ref{fig1}(b), respectively.  Close to depletion, these thin disk-shaped single QDs are prone to split into two parallel-coupled QDs, a splitting we
attribute to an uneven distribution of surface charges. By adjusting the three gate potentials, the interdot tunnel coupling ($t$)  between the first orbitals in the two parallelcoupled QDs can be tuned by an order of magnitude, from weak tunnel coupling to strong tunnel coupling, to finally
form one single QD. More details on the mechanism behind the DQD formation and tuning of the interdot tunnel coupling is provided in Nilsson \textit{et al.} \cite{Nilsson2017}, whereas in this
Letter we focus on the excited-state spectrum and tunability of the spin properties in the 1$e$-2$e$ regime. Figure~\ref{fig1}(c) shows a honeycomb charge stability diagram recorded in the
weak interdot tunnel coupling regime ($t=$~0.45~meV), where the electron population of the DQD can be controlled starting from a fully depleted system.

\begin{figure}[b]
\centering
\includegraphics[width=\columnwidth]{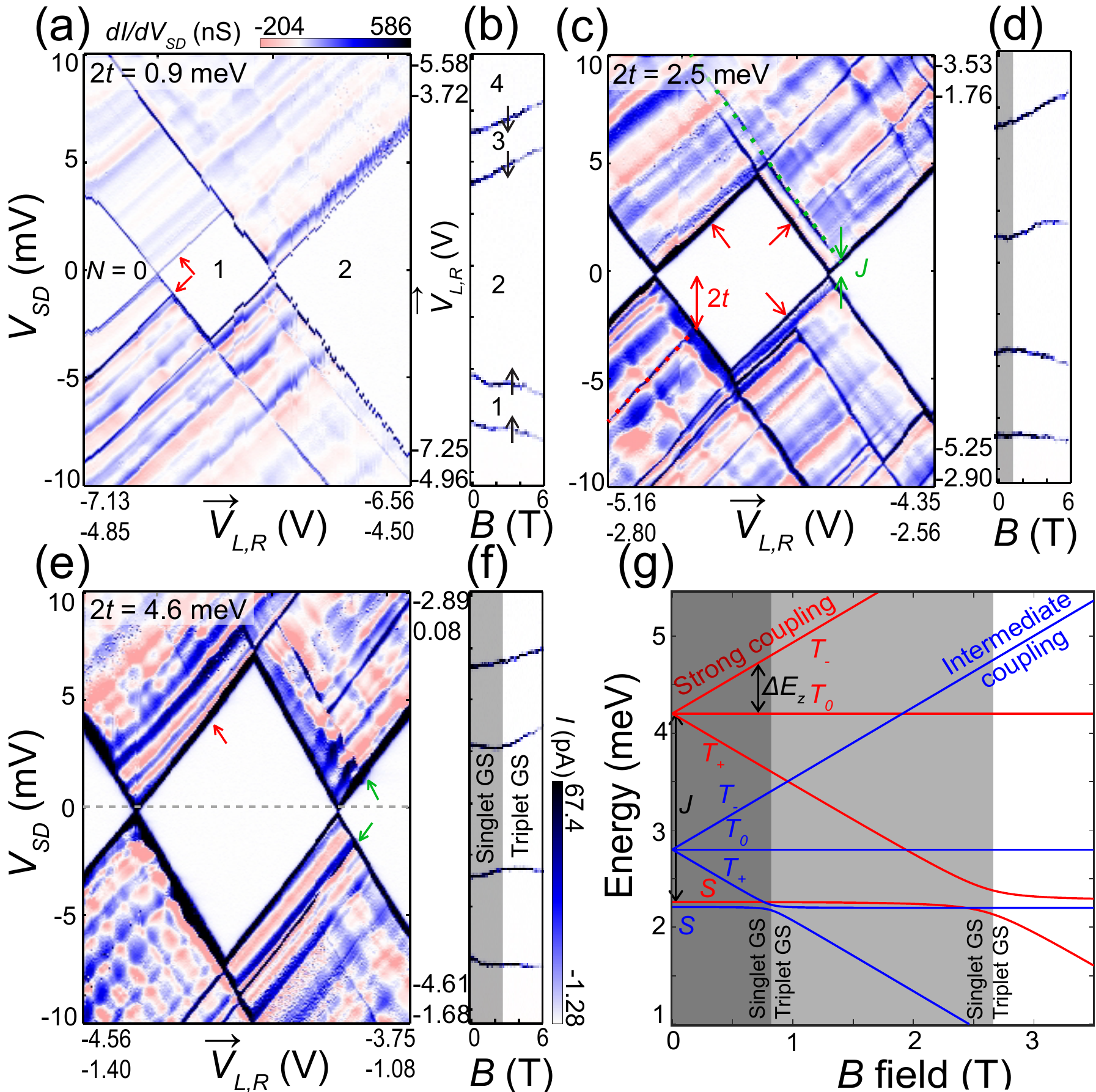}
\caption{(Color online) (a), (c), (e) Coulomb charge stability diagrams for cooldown I, recorded along the side-gate vector $\vec{V}_{L,R}$ crossing the two triple points indicated by the arrow in  Fig.~\ref{fig1}(c). The backgate potential decreases along (a),(c),(e) $\lbrace V_{BG}=0.00,$ [same as Fig.~\ref{fig1}(c)] $ -0.75, -1.25$~V$\rbrace$ effectively increasing the interdot tunnel coupling ($2t$~= 0.9, 2.5, 4.6 $\pm$0.1 meV) and the singlet-triplet energy split ($J$~= 0, 0.6, 1.8 $\pm$0.1 meV). $N$ denotes the total electron population of the DQDs and the arrows indicate the conductance lines associated with transport via the antibonding and triplet states. (b),(d),(f) $B$-field  evolution of the ground states of the first four electrons recorded at $V_{SD} = 20 \mu$V along the side-gate vector illustrated by the dashed line in Fig.~\ref{fig2}(e). Here, the shaded areas indicate where the singlet is ground state, and the arrows indicate the spin direction. (g) Modeled energy of $S$(1,1) and $T$(1,1) states as a function of the $B$ field for intermediate (blue) and strong (red) interdot tunnel coupling; parameters extracted from measurements in the back-gate regimes displayed  areas indicate the singlet-ground-state regimes.
}
\label{fig2}
\end{figure}

Figures~\ref{fig2}(a), \ref{fig2}(c) and \ref{fig2}(e) show Coulomb charge stability diagrams recorded at different back-gate voltages and along the side-gate vector running through the first two
triple points indicated by the arrow in Fig.~\ref{fig1}(c). Along this gate vector, the energies of the two first single-QD orbitals are aligned and the interdot tunnel coupling ($t$), parametrizing the hybridization strength, is extracted from the distance between the conductance lines involving the
1$e$ ground state and first excited state, which in the artificial molecule picture correspond to the bonding ($B$) and antibonding ($AB$) orbitals [see Fig.~\ref{fig2}(c)]  \cite{vanderWiel2002,Hanson2007}. By decreasing $V_{BG}$ (or increasing, not shown here), $t$ is tuned to larger values as a result of the increased overlap of the single-QDs orbitals  \cite{Nilsson2017}.

Looking at the transport via the 2$e$ states, the exchange energy ($J$), defined as the energy difference between the ground-state singlet $S$(1,1) and the first excited-state triplet
$T$(1,1), increases with increasing $t$ as expected \cite{Tokura2009}. We emphasize that the single-QD orbital energy spacings and intradot charging energies are on a much larger energy scale and thus $T$(2,0)/$T$(0,2) and $S$(2,0)/$S$(0,2) do not enter the transport window. From here on,  $S$ and $T$ denote $S$(1,1) and $T$(1,1), respectively.

When an external $B$ field is applied, the $T$ states Zeeman split into three separate states ($T_+$, $T_0$,  $T_-$). 
The $T_+$ state, where both spins are aligned parallel to the $B$ field, decreases in energy by the Zeeman energy $\Delta E_z = |g^*|\mu_B B$ ($\mu_B$  is the Bohr magneton)  and aligns with $S$ when $J_{B=0}=\Delta E_z$,
assuming a linear energy dependence on the $B$ field. 

We model the system as a DQD with a single orbital in each dot  \cite{Stepanenko2012}. Values for  $t$, intradot and interdot charging energies extracted from the experimental results are used as input parameters. Furthermore,  we assume a constant $g^*$ factor estimated from $J_{B=0}=|g^*|\mu_B B_{\Delta}$. Here, $B_{\Delta}$ is the $B$ field at the minimum energy separation of the $S$ and $T_+$ states.  We find that the exchange integral ($V_x$) gives a non-negligible contribution to $J$ for large $t$, for example $V_x$~=~0.258~meV  for $t$~=~2.3~meV (see the Supplemental Material for details on the modeling).

\begin{figure}[b]
\centering
\includegraphics[width=\columnwidth]{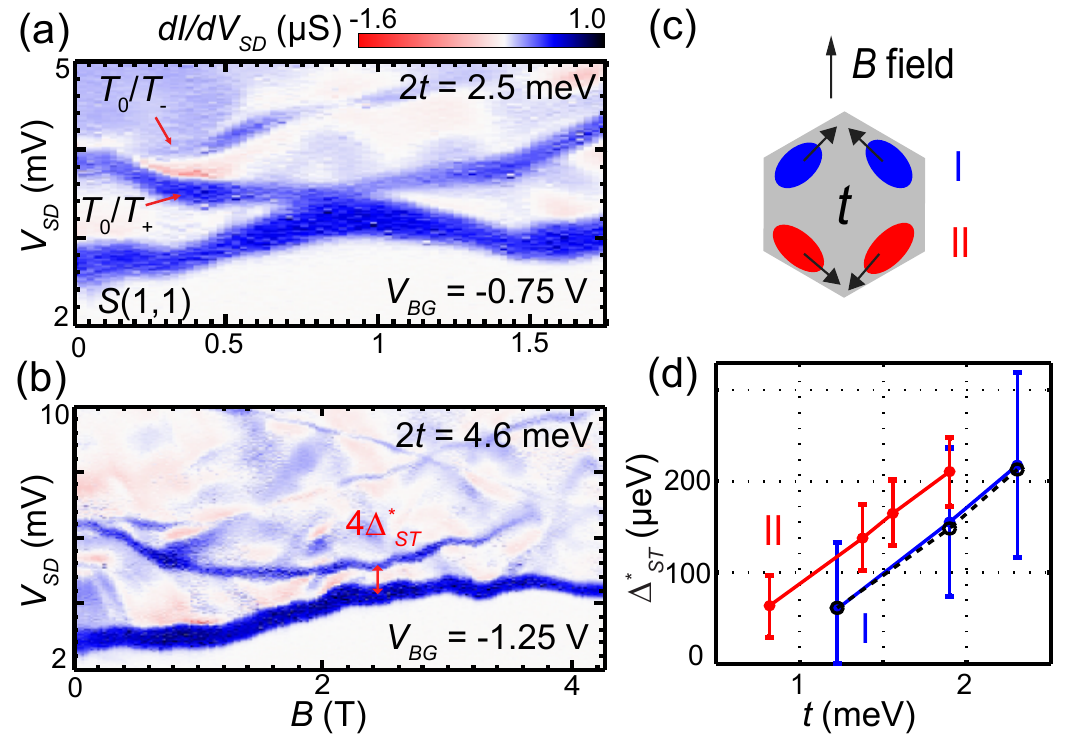}
\caption{$B$-field evolution of transitions involving $S$ and $T$ states recorded during cooldown I at a fixed gate-potential configuration; (a),(b) same $V_{BG}$ as in Figs.~\ref{fig2}(c) and \ref{fig2}(e), respectively.
(a) The 2$e$ states involved in the transitions are indicated. (d) Magnitude of $S/T$ anticrossing $\Delta _{ST}^*$ as a function of interdot tunneling coupling $t$. Blue and red traces correspond to data recorded at two different cooldowns, I and II, respectively. Here, the overlap between the spatial distribution of the electrons in the two dots, parametrized by $t$, is increased by decreasing (blue)
[increasing (red)] $V_{BG}$, as indicated in (c). The error bars reflect the width of the conductance lines. The black dashed trace correspond to results from modeling, where $\Delta _{ST}^*$ has been fitted to data set I by tuning $\alpha$ in the SO-tunneling term $t_{SO}=\alpha t$;  a constant $\alpha$ = 0.08 is used for all $t$ values. }
\label{fig3}
\end{figure}

Figure~\ref{fig2}(g) shows the modeled $B$-field dependence of the $S$ and $T$ states in both the intermediate (blue) and strong (red) tunnel coupling regime, corresponding to
parameters extracted from measurements displayed in Figs.~\ref{fig2}(c) and \ref{fig2}(e), respectively. SOI is included in the model as a spin-nonconserving parameter ($t_{SO} = \alpha t$),  coupling the $S$ and $T_+$ (and $T_-$) states, resulting in an anticrossing near the transition point. states, resulting in an anticrossing near the transition point. For now, we neglect this coupling in the discussion and address the anticrossing further in conjunction with  Fig.~\ref{fig3}.  The hyperfine coupling of electron and nuclear spins is assumed to be small compared to the SOI \cite{Nadj-Perge2010}, and is neglected. 
As $J$ increases with $t$, it follows that $B_{\Delta}$ required to align the $S$ and $T_+$ states also increase with $t$. 
 
The singlet-to-triplet ground-state transition is seen in the $B$-field evolution of the ground state of the first four electron states, probed in Figs.~\ref{fig2}(b), \ref{fig2}(d) and \ref{fig2}(f). The vertical axes are here aligned with the (0,0)-(1,1)-(2,2) triple-points and  $V_{BG}$ values are the same as in Figs.~\ref{fig2}(a), \ref{fig2}(c) and \ref{fig2}(e). At small $t$ and $J$ [Fig.~\ref{fig2}(b)], the negative slope of the two lowest states shows that the ground state is a triplet. Here, the resolution in the $B$ field is not sufficiently high to capture the region where the singlet is ground state. When $t$ and $J$ increase [Figs.~\ref{fig2}(c) and \ref{fig2}(e)], the size of the singlet-ground state region (shaded) increases just as predicted by the model in Fig.~\ref{fig2}(g).

Now we discuss the SOI-induced hybridization of the $S$ and $T$ states, resulting in the 2$e$ ground and excited states being linear combinations of the unperturbed $S$ and $T$
states. The $S$ and $T$ states reach maximum mixing at the singlet- or triplet-ground-state transition point. $T_+$ dominates the contribution to the mixing; thus, the 2$e$ ground and excited states can be written as
\begin{equation}
 GS(1,1) \approx \beta(J) \vert S \rangle - \gamma(J) \vert T_+ \rangle 
\end{equation}
\begin{equation}
 ES(1,1) \approx \gamma(J)\vert S \rangle + \beta(J)\vert T_+ \rangle 
\end{equation}
where $\beta(J)$ and $\gamma(J)$ are $J(B)$ dependent functions.
In the vicinity of the singlet-to-triplet ground-state transition point,  the hybridization give rise to an anticrossing with a magnitude $2\Delta_{ST}^*$  which depends on $t$. Going back to the experimental data, Figs.~\ref{fig3}(a) and \ref{fig3}(b) show the 2$e$-transitions as a function of $B$ field in the intermediate and strong coupling regime. It is clear that both the magnitude of the anticrossing and the $B$ field at the transition point are modulated with $t$. Values forr $\Delta_{ST}^*$ as a function of $t$ were recorded in two separate cooldowns [Fig.~\ref{fig3}(d)] where a stronger tunnel coupling  was obtained by either  applying more negative [blue trace, I], or more positive [red trace, II] $V_{BG}$, as indicated in Fig.~\ref{fig3}(c). We note that the values approach the  $\Delta_{ST}^*$ = 230 $\mu$eV reported
for a gate-defined single InAs WZ QD in the 2$e$ regime \cite{Fasth2005}. Also, moderate gate tuning of the SOI-induced anticrossing  ($\Delta_{ST}^*$ = 50-150 $\mu$eV) for unknown orbitals in single InAs self-assembled QDs has been reported  \cite{Kanai2011}.
In the model, the $\Delta _{ST}^*$ is fitted (dashed trace) by tuning  $\alpha$. With $\alpha= 0.08$ we perform an upper estimate of the spinorbit length using $l_{SO} \sim \frac{t}{t_{SO}}l_{dot} = \frac{1}{\alpha}l_{dot}\approx 400$~nm ($l_{dot}\approx34$~nm is estimated to be half the nanowire diameter), which agrees with previously reported values for InAs nanowires \cite{Nadj-Perge2010,Fasth2007}.

\begin{figure*}
\centering
\includegraphics[width=\textwidth]{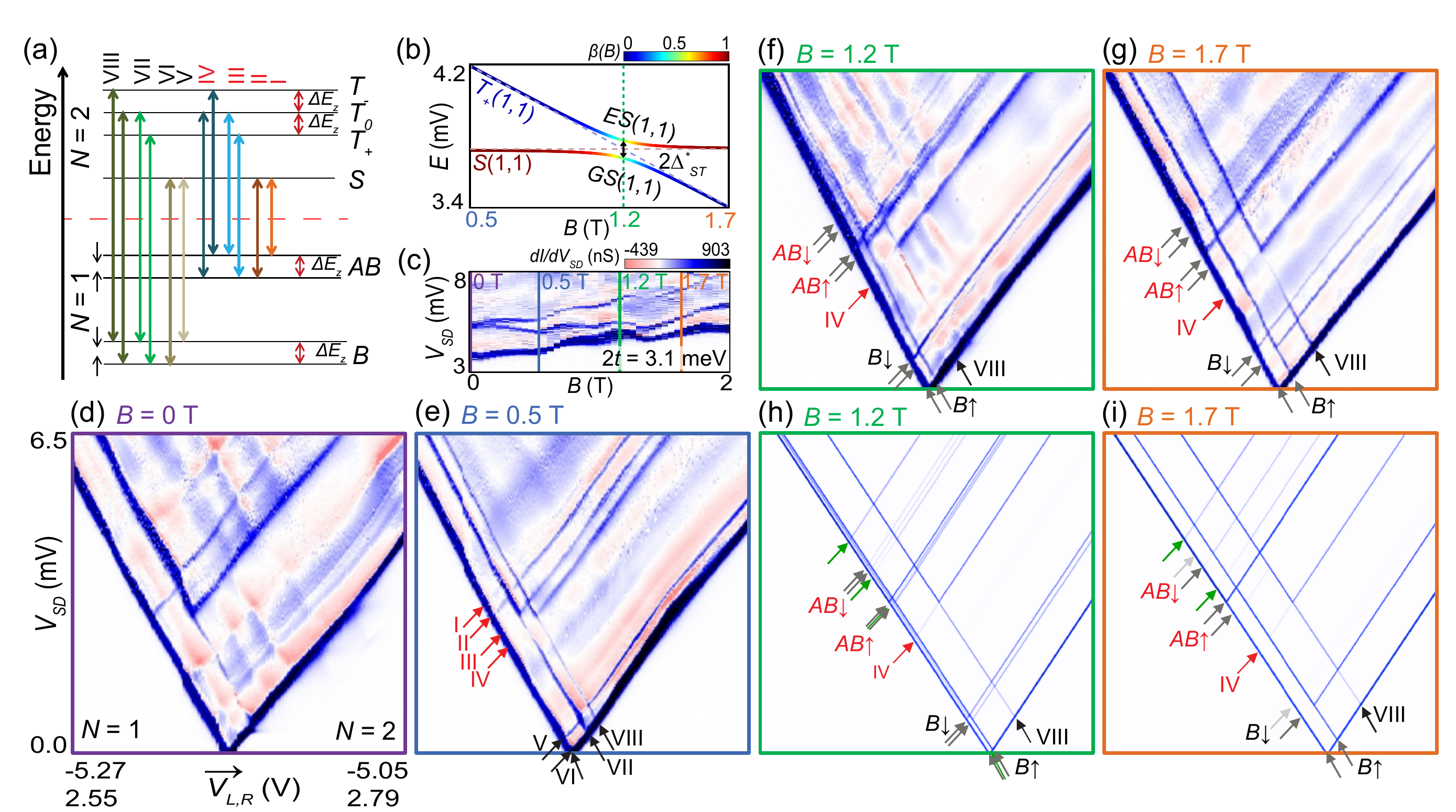}
\caption{
Sketch of the energy levels at a finite external $B$ field and possible transitions between these levels for the 1$e$ and 2$e$ states
(adapted from \cite{Hanson2007}). Here, we assume spin-conserving first-order tunneling processes with no SOI-assisted mixing between states. (b) Modeled energy of the $GS$(1,1) and $ES$(1,1), and the unperturbed $S$ and $T_+$ as a function of magnetic field, where the fraction of $S$ ($T_+$) in $GS$(1,1) [$ES$(1,1)] is indicated by the color. 
(c) Overview of the $B$-field evolution of the 1$e$-2$e$ transitions (cooldown III); the color markings indicate the external $B$ fields at which the Coulomb charge stability diagrams (d)-(g), enlarged on the transitions between 1$e$ and 2$e$ states, are recorded.
(e) Arrows and roman numerals indicate the transitions illustrated in (a); red and black symbolize transitions involving antibonding and bonding 1$e$ states, respectively. 
 (f),(g) The double arrows next to the notation of the 1$e$ states indicate conductance lines involving the hybridized states $GS$(1,1) and $ES$(1,1). (h),(i) Modeling using a $B$ field corresponding to (f) and (g), respectively, where the transitions seen in the experimental data are indicated. Additional lines corresponding to transitions
involving $T_0$ and 1$e$ - 0$e$ are indicated by green arrows.}
\label{fig4}
\end{figure*}

In Fig.~\ref{fig4}, we take a closer look at the gate-dependent transport involving 1$e$ and 2$e$ ground and excited states for the four different $B$ fields indicated in Fig~\ref{fig4}(b). Owing to the large $|g^*|$ factors ($\sim$10) and the long lifetime of excited states, we see clearly separated and very sharp conductance lines associated with transitions starting from excited states. As a result, we can resolve all conductance lines predicted by a simple model, including possible first-order tunneling
transitions between 1$e$ and 2$e$ states at low $B$ fields, assuming spin conservation[Fig.~\ref{fig4}(a)].
Arrows and roman numerals associate the transitions in Fig.~\ref{fig4}(a) to  the corresponding conductance lines in Figs~\ref{fig4}(e)-(g).

As the $B$ field increases, going from Fig.~\ref{fig4}(d) to Fig.~\ref{fig4}(e),  the conductance lines corresponding to transport via bonding ($B$), antibonding ($AB$), and $T$ states Zeeman split. Assuming that $\Delta E_z$ for these splits are equal and no
spin-flip processes occur, the four different transitions between $B$ and $T$ states pairwise become possible at the same energy and are mapped to two different transition
energies; thus, only two conductance lines appear in Fig.~\ref{fig4}(e). The same argument holds for transitions between $AB$ and $T$ states.

When increasing the $B$ field further, as in Fig.~\ref{fig4}(f), the anticrossing of $GS$(1,1) and $ES$(1,1) is reflected in an apparent pairing of the conductance lines previously associated with transitions involving unperturbed $S$ and
$T_+$ states. Although some lines are weak, we can discern all lines, except for the excited-state-to-excited-state transitions $B\downarrow$ - $T_0$ and $AB\downarrow$ + $T_0$, predicted when a SOI-induced mixing of $S$ and $T_+$ is included in our simple model depicted in Fig.~\ref{fig4}(a). Assuming that $B$~=~1.2~T  is the
field where the unperturbed levels cross, the energy separation between the pairs correspond to $2 \Delta_{ST}^*$. Here,
double arrows indicate transitions from the noted 1$e$ state to $GS$(1,1) and $ES$(1,1) states. This paired conductance-line
pattern is well reproduced by the modeling [see Fig.~\ref{fig4}(h)]. The origin of the lines in the model is further discussed in
the Supplemental Material.
 
Although the mixing of $S$ and $T_{+}$ is weaker at larger $B$ field (larger detuning) [Fig.~\ref{fig4}(g) and \ref{fig4}(i)], the hybridization is strong enough for the spin-nonconserving transitions to be visible, such as the upper line in the $B\downarrow$ ($AB\downarrow$)  pair, and thus the pairing of all lines persists. Here, the increased
separation of the paired lines is a result of the detuning of the  (1,1) states as shown in Fig.~\ref{fig4}(b).

In conclusion, we tune the interdot tunnel coupling of parallel-coupled DQDs formed in InAs nanowires. This directly controls the $B$-field-dependent singlet-to-triplet transition of the two-electron ground state and also the magnitude of the SOI-induced anticrossing energy. Owing to long life times, large $|g^*|$ factor and strong SOI, the $B$-field evolution of two-electron states is clearly resolved. The experimental results are reproduced with a simple DQD
model Hamiltonian that includes a constant SOI term. As a
next step, this system could be used for studies of manybody correlated transport such as orbital- \cite{JarilloHerrero2005} and spin-Kondo effects \cite{Krychowski2016} and Cooper-pair splitting \cite{Deacon2015,Baba2015}.

\begin{acknowledgments}
The authors thank M. Hell, A. Burke and A. Kvennefors for support and discussions. This work was carried out with financial support from NanoLund, the Swedish Research Council (VR), the Swedish Foundation for Strategic Research (SSF), the Crafoord Foundation,  and the Knut and Alice Wallenberg Foundation (KAW).
\end{acknowledgments}

\bibliography{Spin_DQD_manuscript2_ref}

\end{document}